\documentstyle[amssymb,aps,multicol]{revtex}

\begin{document}
\draft
\title{Principles of statistical mechanics of random networks }
\author{S.N. Dorogovtsev$^{a,b,*}$, J.F.F. Mendes$^{c,\dagger }$, and A.N. Samukhin$%
^{a,b,\ddagger }$}
\address{$^a$ Departamento de F\'\i sica and Centro de F\'\i sica do Porto,
Faculdade Ci\^{e}ncias, Universidade do Porto,\\ Rua do Campo Alegre 687,
4169-007 Porto, Portugal\\
$^b$ A.F. Ioffe Physico-Technical Institute, 194021 St. Petersburg, Russia\\
$^c$Departamento de F\'\i sica, Universidade de Aveiro, Campus Universit\'{a}rio
de Santiago, 3810-193 Aveiro, Portugal}
\maketitle

\begin{abstract}
We develop a statistical mechanics approach for random networks with uncorrelated vertices. We
construct equilibrium statistical ensembles of such networks and obtain their
partition functions and main characteristics. We find simple dynamical
construction procedures that produce equilibrium uncorrelated random graphs with an arbitrary degree distribution. In particular, we show 
that in equilibrium uncorrelated networks, fat-tailed degree distributions may exist only
starting from some critical average number of connections of a vertex, in a
phase with a condensate of edges.
\end{abstract}

\pacs{{\em PACS:} 05.10.-a, 05.40.-a, 05.50.+q, 87.18.Sn} 

\noindent 
{\footnotesize {\em Key words:} Random graphs, Statistical ensembles, Zero-dimensional field theory, Backgammon model}


\section{Introduction}
\label{intro}

Quite recently, it has been realized that the study of the structure of networks,
which was formerly ``a private domain'' of mathematical graph theory \cite
{er59,bb85,mr95}, is actually a field of statistical physics \cite
{ba99,w99,s01,ab01a,dm01c,dmbook03}. Most achievements of physicists in the field of
random networks \cite{remark0} (structural and topological aspects) are
empirical findings and simple ideas, which were demonstrated by using very
particular models. One of main questions that arise is: What is the nature
of complex, non-Poisson 
degree distributions, which were observed in many real
networks? (Degree of a vertex is the total number of its connections.)
However, few efforts were made to develop a general statistical theory
of networks (see Refs. \cite{bck01,k01,bl02,bk02}). Without such a theory, the above question cannot be answered. 
Furthermore, the structure of the statistical theory of random networks can be related to a zero-dimensional field theory and to a mean-field description of simplicial gravity \cite{bbj99}. A condensation phase transition, which occurs in equilibrium networks, is close to that which occurs in branched polymers \cite{adj90,aj95}.  

In this paper we focus on equilibrium network ensembles, which are less
studied. We construct statistical ensembles of random uncorrelated networks in a natural
way and establish a one-to-one correspondence between them and known
construction procedures. One of conclusions is that, in equilibrium
networks, fat-tailed degree distributions (in particular, power-law ones)
are possible only starting from some critical value of the average degree.
Above this critical point, a finite fraction of edges are in a ``condensed
state'', i.e. are attached to an infinitely small fraction of vertices.
This situation for equilibrium networks contrasts with that for growing
ones, were fat-tailed degree distributions are realized over a wide range of
a control parameter without any condensate.

The keystone of network science are construction procedures. Erd\"{o}s and
R\'{e}nyi constructed ensembles of random graphs with a Poisson degree
distribution by adding edges at random to a fixed number of vertices \cite
{er59}. When the total number of edges $L$ is fixed, this provides a
canonical ensemble. When one fixes the probability that two vertices are
connected, the procedure produces a grand canonical ensemble.

To obtain equilibrium random graphs with an arbitrary degree distribution $%
\Pi \left( q\right) $, a different statistical ensemble was introduced \cite
{mr95} (see also Ref. \cite{nsw00}). Roughly speaking, these are graphs,
maximally random under the restriction that their degree distribution is
equal to a given $\Pi \left( q\right) $ (see below). Here we demonstrate
that this ``static'' construction produces a microcanonical ensemble, and
construct equivalent (in the thermodynamic limit, i.e. $N\to \infty $) canonical
and grand canonical ensembles as limiting equilibrium states of simple
dynamical processes.

In statistical mechanics, equilibrium ensembles arise as infinite-time
limiting distributions of some ergodic dynamics. Here we present naturally
looking graph evolution models, using the generalization of the concept of
``preferential linking'', which was introduced in \cite{ba99}. We consider
two kinds of a random network evolving under the mechanism of preferential
linking and displaying ergodic behavior. The latter means that an evolving
ensemble finally becomes equilibrium, and final statistical weights for the
complete set of graphs of the ensemble are independent of time and initial
state. The specific rule of preferential linking that we use allows us to
construct equilibrium statistical ensembles with an arbitrary $\Pi \left(
q\right) $.

The paper is organized as follows. In Section \ref{defs} we introduce the
main notions of random graph theory. The next Section \ref{undirect} is a
key one: we establish a connection between ergodic evolution and statistical
ensembles for undirected graphs. In Section \ref{direct}
these results are generalized to the case of directed graphs. Section \ref
{fat} is devoted to networks with a {\em fat-tailed} degree distribution, decaying slower than exponential. The results are discussed in
the last Section \ref{concl}. Technical details are described in three
Appendices.

\section{Definitions and main notions}
\label{defs}

{\em A graph} $g$ is a set of $N$ {\em vertices} connected by $L$ {\em edges}%
, directed or undirected. It may be displayed as a set of points, with some
pairs connected by one or more lines, with or without arrows. For analytic
purposes, a graph is represented by an $N\times N$ {\em adjacency matrix} $\widehat{g%
}$, whose elements $g_{ij}$ are numbers of edges connecting vertices $i$
and $j$.

An {\em undirected} graph (with undirected edges), is represented by a
symmetric adjacency matrix, $g_{ij}=g_{ji}$. In this case it is convenient
to set diagonal elements $g_{ii}$ to be equal to twice the number of
unit-length loops. Then, the adjacency matrix $\widehat{g}^{(1)}$ of an
undirected graph $G_1$, obtained from a directed graph $G$ with an adjacency
matrix $\widehat{g}$ by replacing all directed edges with undirected ones,
is simply $g_{ij}^{(1)}=g_{ij}+g_{ji}$.

{\em Mayer's graphs} are the ones without multiple connections and
one-vertex loops. Their adjacency matrix elements satisfy the conditions: $%
g_{ij}^2=g_{ij}$, $g_{ii}=0$.

{\em Vertex in-degree} $r_i$ of a vertex $i$ in a directed graph is a number
of incoming edges of the vertex $i$, $r_i=\sum_jg_{ij}$. Similarly, {\em %
out-degree} $s_i$ is a number of edges, outgoing from the vertex $i$, $%
s_i=\sum_jg_{ji}$. {\em Vertex degree }for undirected graph is $%
q_i=\sum_jg_{ij}=\sum_jg_{ji}$.

{\em A statistical ensemble of graphs} is defined by choosing of a set $G$
of graphs and a rule that associates some {\em statistical weight}
(unnormalized measure) $P(g)>0$ with any graph $g\in G$. Then the ensemble
average of any quantity $A\left( g\right) $ that depends on properties of a
graph is $\left\langle A\right\rangle =Z^{-1}\sum_{g\in G}A\left( g\right)
P\left( g\right) $, where $Z$ is a partition function, $Z\equiv \sum_{g\in
G}P\left( g\right) $. For instance, let $A\left( g\right) $ be the total
number of vertices of in-degree $r$ and out-degree $s$: 
\begin{equation}
N\left( r,s;g\right) =\sum_{i=1}^N\delta \left[ r_i\left( g\right) -r\right]
\delta \left[ s_i\left( g\right) -s\right] \,.  \label{5}
\end{equation}
Here $N$ is the total number of vertices in the graph $g$ (we consider only
ensembles with a fixed number of vertices). The probability that a randomly
chosen vertex has in-degree $r$ and out-degree $s$ (a degree distribution)
is 
\begin{equation}
\Pi \left( r,s\right) \equiv \frac{\left\langle N\left( r,s\right)
\right\rangle }N=\frac 1N\left\langle \sum_{i=1}^N\delta \left[ r_i\left(
g\right) -r\right] \delta \left[ s_i\left( g\right) -s\right] \right\rangle 
.
\label{10}
\end{equation} 
For an undirected graph, one can define the number of vertices with a given
degree $q$ 
\begin{equation}
N\left( q\right) =\sum_{i=1}^N\delta \left[ q_i\left( g\right) -q\right] 
\,,
\label{15}
\end{equation}
and a degree distribution 
\begin{equation}
\Pi \left( q\right) =\frac{\left\langle N\left( q\right) \right\rangle }N=%
\frac 1N\left\langle \sum_{i=1}^N\delta \left[ q_i\left( g\right) -q\right]
\right\rangle 
\,.  
\label{17}
\end{equation}

In this paper we consider statistical ensembles with non-Mayer's graphs
allowed. The advantage of this assumption is that one can associate a
statistical weight with a graph by using the same rules, as for the
contribution of the corresponding Feynman diagram in an appropriately chosen
zero-dimensional field theory (see \cite{bck01,k01}).

Note that we consider labeled graphs. That is, two graphs, $g$ and $%
g^{\prime }$, which differ only by numeration of vertices, are treated as
different ones.

\section{Evolution of graphs and statistical ensembles: Undirected graphs}
\label{undirect}

In this section we discuss how ensembles of undirected graphs arise as a
result of the network evolution. For simplicity, we restrict ourselves to
undirected graphs---generalization to the case of directed ones is
presented in the next section. We define the statistical weights of the
canonical and grand canonical ensembles of random networks as a limiting 
{\em equilibrium} distribution of a process, during which one graph $g\in G$
of the ensemble transforms to another graph $g^{\prime }\in G$ with
probability $W\left( g^{\prime },g\right) dt$. The statistical weights $%
P\left( g,t\right) $ evolve according to the master equation 
\begin{equation}
\partial _tP\left( g,t\right) =\sum_{g^{\prime }\in G}\left[ W\left(
g,g^{\prime }\right) P\left( g^{\prime },t\right) -W\left( g^{\prime
},g\right) P\left( g,t\right) \right]  
\, .
\label{20}
\end{equation}
$\,$An equilibrium ensemble is a stationary one. $P(g,t)=P(g)$ is
independent of $t$, where statistical weights are determined by the detailed
balance condition (absence of ``currents''): 
\begin{equation}
W(g,g^{\prime })P(g^{\prime })=W(g^{\prime },g)P(g)  
\, . 
\label{22}
\end{equation}
This equilibrium ensemble exists if and only if the set of ``hopping rates'' 
$W(g,g^{\prime })$ satisfies two conditions:  
(i) For any pair of graphs $%
g,g^{\prime }\in G$, there exists a sequence of graphs $g_1,g_2,\dots g_n\in
G$ such that 
\begin{equation}
W(g^{\prime },g_n)W(g_n,g_{n-1})\dots W(g_2,g_1)W(g_1,g)\neq 0  
\, .
\label{24}
\end{equation}
(ii) For any sequence of graphs $g_1,g_2,\dots g_n\in G$, the equality: 
\begin{equation}
W(g_1,g_2)W(g_2,g_3)\dots
W(g_{n-1},g_n)W(g_n,g_1)=W(g_1,g_n)W(g_n,g_{n-1})\dots
W(g_3,g_2)W(g_2,g_1)
\,,  
\label{25}
\end{equation}
is valid. 

These conditions ensure that (a) ascribing an arbitrary statistical weight
to some graph, one can obtain statistical weights of all other graphs up to
a constant multiple, and (b) this definition is unambiguous: the weights are
independent of the ways connecting initial graph with all the other ones. To
satisfy condition (ii), it is sufficient to assume the factorization:
\begin{equation}
W(g^{\prime },g)=V_f(g^{\prime })V_i(g)\,.  \label{30}
\end{equation}
Our dynamical constructions, which are presented below, satisfy this
condition. We use simple natural assumptions about the evolution rates $%
W\left( g^{\prime },g\right) $, but our choice is not unique (e.g., see a
``Metropolis algorithm'' from Refs. \cite{bck01,k01}).

We consider the following equilibrium statistical ensembles of graphs {\em %
with a fixed total number of vertices} $N$. A statistical ensemble is a
set of graphs $G$ plus rules that determine statistical weighs $P\left(
g\right) $ for all graphs $g\in G$.

1. A microcanonical ensemble \cite{mr95}.

\begin{itemize}

\item[({\em set})] Let $N\left( q\right) $ be a sequence of non-negative
integers such that $0<\sum_qN\left( q\right) =N<\infty $. $G_{MC}$ is the set
of all graphs of size $N$, for which number of vertices of degree $q$ is
equal to $N\left( q\right) $.

\item[({\em rule})] To each graph $g\in G_{MC}$ ascribe the weight  
\begin{equation}
P_{MC}\left( g\right) =N^{-L}\prod_{i=1}^N\frac{q_i!}{g_{ii}!!}%
\prod_{j<k=1}^N\frac 1{g_{jk}!}
\,.  
\label{35}
\end{equation} 

\end{itemize}

This is a ``static'' construction. These statistical weights follow from
pure combinatorics. They are just the number of possible ways to obtain a
given graph $g\in G_{MC}$ by connecting together $N$ vertices with degrees $%
q_1,q_2,\dots q_N$ (see proof in Appendix \ref{combinatorics}). The multiple 
$N^{-L}$ is introduced to ensure the extensiveness of the ``free energy'', $\ln
Z_{MC}$. Eq. (\ref{35}) implies, that edges in the graph are
distinguishable. Note that if only Mayer graphs are allowed, all graphs in
this ensemble have equal weights. In the thermodynamic limit \cite{remark0a}%
, the microcanonical ensemble is described by a sequence of values $\left\{
\Pi \left( q\right) \right\} $ or, which is the same, $\left\{ N\left(
q\right) \right\} $ (in particular, this includes the mean degree $\bar{q}%
\leftarrow 2L/N$). 

To construct canonical and grand canonical ensembles we use the processes of
rewiring \cite{ab00a} or of deletion/creation of edges \cite{dm00}, and the
idea of preferential linking \cite{ba99,krl00,dms00}. We assume that the probability
that an edge becomes attached to a vertex $i$ depends only on the degree $%
q_i $ of this vertex. This probability is determined by some preference
function $f\left( q\right) $.

2. A canonical ensemble.

\begin{itemize}

\item[({\em set})] The set $G_C$ consists of all graphs with $N$ vertices
and $L$ edges.

\item[({\em rule})] At each step of the evolution, one of the ends of a randomly
chosen edge is rewired to a preferentially chosen vertex $k$. Let the rate
of this process be $f\left( q_k\right) $ \cite{remark0b}. The limiting
stationary statistical weights give $P_C\left( g\right) $. In the
thermodynamic limit, the canonical ensemble is described by $\left\{ f\left(
q\right) \right\} $ and $\bar{q}\leftarrow 2L/N$. Note that the
multiplication of $f\left( q\right) $ by a constant, $f\left( q\right) \to
Cf\left( q\right) $, is simply the rescaling of time, $t\to t/C$. It does
not influence equilibrium properties.
\end{itemize}

3. A grand canonical ensemble.

\begin{itemize}
\item[({\em set})] 
The set $G_{GC}$ consists of all graphs with any number of edges  
and a fixed number of vertices, $N$.

\item[({\em rule})] There are two parallel processes in this case: edges are
deleted and emerge permanently. Randomly chosen edges are deleted at a rate $%
\lambda N$ ($\lambda $ is the inverse lifetime of an edge, $\lambda $ is
fixed as $N\to \infty $, i.e., in the thermodynamic limit). Edges between
vertices $i$ and $j$ emerge at a rate $f\left( q_i\right) f\left(
q_j\right) $. To ensure the correspondence with the canonical ensemble, let
the deletion rate of tadpoles be $2\lambda N$. 
\end{itemize} 
In the thermodynamic limit, 
the grand canonical ensemble is described by $\left\{ f\left( q\right)
\right\} $ and $\lambda $.

Let us obtain, for example, statistical weights for the canonical ensemble.
Let an edge $\left( i,j\right) $ of a graph $g$ be rewired to $\left(
i,k\right) $ in a graph $g^{\prime }$. We have the following balance
equation for the statistical weights of these two graphs:
\begin{equation}
g_{ik}^{\prime }f\left( q_j^{\prime }\right) P_C\left( g^{\prime }\right)
=g_{ij}f\left( q_k\right) P_C\left( g\right) \,.  \label{40}
\end{equation}
Here quantities with a prime mark are referred to the graph $g^{\prime }$, $%
q_j^{\prime }=q_j-1$, $g_{ik}^{\prime }=g_{ik}+1+\delta _{ik}$ (adding a
tadpole increases $g_{ii}$ by two). The multiple $g_{ij}$ is present, because
rewiring any of $\left( i,j\right) $ edges gives the same result. One can
look for the solution in the form:
\begin{equation}
P_C\left( g\right) =N^{-L}\prod_{i=1}^Np\left( q_i\right) \chi _d\left(
g_{ii}\right) \prod_{j<k=1}^N\chi \left( g_{jk}\right) \,,  \label{50}
\end{equation}
where $p$, $\chi $ and $\chi _d$ are some functions of an integer argument.
Substituting Eq. (\ref{50}) into Eq. (\ref{40}), we obtain at $i\ne j$, $%
i\ne k$: $p\left( q+1\right) =f\left( q\right) p\left( q\right) $, $\chi
\left( g+1\right) =\chi \left( g\right) /\left( g+1\right) $. Setting $i=j$
or $i=k$, we get: $\chi _d\left( g+2\right) =\chi _d\left( g\right) /\left(
g+2\right) $. The constant multiple $N^{-L}$ is introduced to ensure the
``free energy'' to be extensive variable, $\ln Z_C\sim N$. Thus we obtain:
\begin{eqnarray}
p\left( q\right) &=&\prod_{r=0}^{q-1}f\left( r\right) \ \ \mbox{for}\ \ q>0\
,\ \ \ p\left( 0\right) =1\,,\;  \nonumber \\
\chi \left( g\right) &=&\frac 1{g!}\,,\;\chi _d\left( g\right) =\frac 1{g!!} 
\, .
\label{60}
\end{eqnarray}
Then we have  
\begin{equation}
P_C\left( g\right) =N^{-L}\prod_{i=1}^N\frac{p\left( q_i\right) }{g_{ii}!!}%
\prod_{j<k=1}^N\frac 1{g_{jk}!}  
\, .
\label{62}
\end{equation}
Comparing Eq. (\ref{62}) with Eq. (\ref{35}), one can see that  
\begin{equation}
P_C\left( g\right) =P_{MC}\left( g\right) \prod_{i=1}^N\frac{p\left(
q_i\right) }{q_i!}\,.  \label{63}
\end{equation}
Analogously, for the grand canonical ensemble we have
\begin{equation}
P_{GC}\left( g\right) =\left( \lambda N\right) ^{-L(g)}\prod_{i=1}^N\frac{%
p\left( q_i\right) }{g_{ii}!!}\prod_{j<k=1}\frac 1{g_{jk}!}\,=\lambda
^{-L\left( g\right) }P_C\left( g\right) ,  \label{65}
\end{equation}
where $p\left( q\right) $ is again given by Eq. (\ref{60}). 
Here $L(g)$ is the number of edges in a graph $g$. 

One can present the statistical weights in a different form. For the
canonical ensemble, one can write
\begin{equation}
P_C\left( g\right) =\prod_{i=1}^N\frac 1{g_{ii}!!}\prod_{j<k=1}\frac 1{%
g_{jk}!}\exp \left[ \sum_{q=0}^\infty N\left( q,g\right) \ln p\left(
q\right) \right] \,,  \label{67}
\end{equation}
The corresponding form for the grand canonical ensemble includes the
additional term $-L\left( g\right) \ln \left( \lambda N\right) $ in the
exponential: 
\begin{equation}
P_{GC}\left( g\right) =\prod_{i=1}^N\frac 1{g_{ii}!!}\prod_{j<k=1}\frac 1{%
g_{jk}!}\exp \left[ -L\left( g\right) \ln \left( \lambda N\right)
+\sum_{q=0}^\infty N\left( q,g\right) \ln p\left( q\right) \right] \,.
\label{68}
\end{equation}

The above constructions are reasonable only if these ensembles are
equivalent in the thermodynamic limit. Here we show that this is the case.
One can see from Eq. (\ref{67}) that statistical weights of graphs with the
same sequences $\left\{ N\left( q,g\right) ,\,q=0,1,2,\dots \right\} $ are
equal. Then the canonical ensemble is equivalent to the microcanonical one
with the same $\Pi (q)=\left\langle N\left( q\right) \right\rangle /N$ if
fluctuations of $N\left( q\right) $ are negligibly small in the
thermodynamic limit: $\left[ \left\langle N^2\left( q\right) \right\rangle
-\left\langle N\left( q\right) \right\rangle ^2\right] /\left\langle N\left(
q\right) \right\rangle ^2\to 0$ as $N\to \infty $. To study these
fluctuations, one may use the following standard relations:
\begin{eqnarray}
&&\frac{\delta \ln Z\left( \left\{ p\left( \tilde{q}\right) \right\} \right) 
}{\delta \ln p\left( q\right) }=\left\langle N\left( q\right) \right\rangle
\,,  \nonumber \\[0.07in]
&&\frac{\delta ^2\ln Z\left( \left\{ p\left( \tilde{q}\right) \right\}
\right) }{\delta \ln p\left( q\right) \delta \ln p\left( q^{\prime }\right) }%
=\left\langle N\left( q\right) N\left( q^{\prime }\right) \right\rangle
-\left\langle N\left( q\right) \right\rangle \left\langle N\left( q^{\prime
}\right) \right\rangle \,,  \label{70}
\end{eqnarray}
which follows from Eq. (\ref{67}) and the definition of the partition
function. Equation (\ref{70}) holds both for the canonical and the grand
canonical ensembles.

Notice that the transition from a microcanonical ensemble to canonical one
is basically the Legendre transform \cite{v98}, where some thermodynamically
conjugated fields are used. In our case, the microcanonical ensemble is
characterized by a sequence of $\left\{ N\left( q\right) \right\} $, and the
conjugated fields are $\left\{ \ln p\left( q\right) \right\} $. In the grand
canonical ensemble, $-\ln \left( \lambda N\right) $ is analogous to a
standard chemical potential or, more precisely, to $\mu /kT$.

The partition function of the grand canonical ensemble is
\begin{equation}
Z_{GC}\left( N,\lambda ,\left\{ p\left( q\right) \right\} \right)
=\sum_{L=0}^\infty \lambda ^{-L}Z_C\left( N,L,\left\{ p\left( q\right)
\right\} \right) \,.  \label{80}
\end{equation}
Let us introduce a zero-dimensional theory of real scalar field $%
x$ with the action \cite{bck01,k01} 
\begin{equation}
S\left( x\right) =-\frac \Lambda 2x^2-\varkappa \Phi \left( x\right) \,,
\label{85}
\end{equation}
where
\begin{equation}
\Phi \left( x\right) =\sum_q\frac{p\left( q\right) }{q!}x^q\,.  \label{100}
\end{equation}
Then the generating functional of this theory can be expanded in the series
of all possible Feynman diagrams, whose contributions coincide with
statistical weights: 
\begin{equation}
Z\left( \Lambda ,\varkappa ,\left\{ p\left( q\right) \right\} \right) =\sqrt{%
\frac \Lambda {2\pi }}\int_{-\infty }^{+\infty }dx\,\exp S\left( x\right)
=\sum_{N=0}^\infty \frac{\left( -\varkappa \right) ^N}{N!}Z_{GC}\left(
N,\Lambda /N,\left\{ p\left( q\right) \right\} \right) 
\,.  
\label{102}
\end{equation}
Then we come to the expression
\begin{equation}
Z_{GC}\left( N,\lambda ,\left\{ p\left( q\right) \right\} \right) =\sqrt{%
\frac{N\lambda }{2\pi }}\int_{-\infty }^{+\infty }dx\,\exp \left( -\frac{%
N\lambda }2x^2\right) \left[ \Phi \left( x\right) \right] ^N 
\, .
 \label{90}
\end{equation}

From Eq. (\ref{80}) it follows that  
\begin{equation}
Z_C\left( N,L,\left\{ p\left( q\right) \right\} \right) =\oint_C\frac{%
d\lambda }{2\pi i}\lambda ^{L-1}Z_{GC}\left( N,\lambda \right) \,,
\label{105}
\end{equation}
where the integration contour $C$ has no singularities outside of it.
Substituting Eq. (\ref{90}) into Eq. (\ref{105}), changing the order of
integration, and calculating the integral over $\lambda $, we have  
\begin{equation}
Z_C\left( N,L,\left\{ p\left( q\right) \right\} \right) =N^{-L}\left(
2L-1\right) !!\oint_c\frac{dx}{2\pi i}x^{-2L-1}\left[ \Phi \left( x\right)
\right] ^N.  \label{107}
\end{equation}
where the contour $c$ encircles the point $x=0$. This derivation is rather
formal, because convergence of the integrals for the generating functional $Z$,
Eq. (\ref{102}), and for the grand canonical partition functions, Eq. (\ref
{90}), depends on the properties of $\Phi \left( x\right) $. These integrals
are well defined and converge if the diagrammatic series for the corresponding
partition functions, or for the generating functional, converge. Note that
the partition function of the grand canonical ensemble does not exist if
either $\ln \Phi \left( x\right) $ is growing faster than $x^2$ at $%
x\rightarrow \infty $, or $\Phi \left( x\right) $ has singularities at the
real axis (series (\ref{100}) has a finite radius of convergence). But for
the canonical ensemble, the partition function does exist for every $\Phi
\left( x\right) $, which is analytic at $x=0$. Indeed, for the canonical
ensemble, the partition function is a sum over a finite set of graphs, while
for the grand canonical ensemble, it is an infinite series, which may
diverge. More detailed derivation of Eqs. (\ref{90}) and (\ref{107}) is
presented in Appendix \ref{diagrams}. Note also that the expression (\ref
{107}) coincides with that for the partition function of the backgammon (``balls
in boxes'') model \cite{bbj99}.

As $N\to \infty $, one can use the saddle point expression:
\begin{equation}
Z_C\left( N,L,\left\{ p\left( q\right) \right\} \right) \rightarrow \left( 
\frac{\bar{q}}{ex_s^2}\right) ^L\left[ \Phi \left( x_s\right) \right] ^N\,,
\label{110}
\end{equation}
where $\bar{q}=2L/N$ is the average vertex degree and the saddle point $x_s$
is given by the equation 
\begin{equation}
\bar{q}=x_s\frac{\Phi ^{\prime }\left( x_s\right) }{\Phi \left( x_s\right) }%
\,.  \label{120}
\end{equation}
We omitted a preexponential saddle-point multiple in Eq. (\ref{110}) as
insignificant in the thermodynamic limit. For grand canonical ensemble we
have:  
\begin{equation}
Z_{GC}\left( N,\lambda ,\left\{ p\left( q\right) \right\} \right) =\exp
\left( -\frac{N\lambda }2x_s^2\right) \left[ \Phi \left( x_s\right) \right]
^N\,,  \label{125}
\end{equation}
\begin{equation}
\lambda x_s=\frac{\Phi ^{\prime }\left( x_s\right) }{\Phi \left( x_s\right) }%
\,.  \label{127}
\end{equation}

From the fact that the logarithm of the partition function of the canonical
ensemble is extensive, $\ln Z_{GC}\sim N$ 
(see Eqs. (\ref{70}) and (\ref{110})), it
follows that $\left\langle N\left( q\right) N\left( q^{\prime }\right)
\right\rangle -\left\langle N\left( q\right) \right\rangle \left\langle
N\left( q^{\prime }\right) \right\rangle ={\cal O}\left( N\right) $, so that
the canonical ensemble is indeed equivalent to the microcanonical one.
Analogously, using the relations 
\begin{eqnarray}
& & 
\left\langle L\right\rangle = 
- \frac{\partial \ln Z_{GC}}{\partial \ln \lambda} 
\,,
\nonumber
\\[5pt]
& &
\left\langle L^2\right\rangle -\left\langle L\right\rangle ^2=
\frac{\partial^2 \ln Z_{GC}}{\partial (\ln \lambda)^2}
= {\cal O}\left( N\right) 
\, , 
\label{128}
\end{eqnarray}
one finds that the fluctuations of the number of edges $L$ in the grand
canonical ensemble disappear in the thermodynamic limit. This demonstrates
the equivalence of the grand canonical and canonical ensembles, if $f\left(
q\right) $ grows not very fast with $q$, which allows the existence of the
grand canonical ensemble. Their parameters are related as: $\lambda
=Lx_s^2/N=\bar{q}x_s^2$.

From Eqs. (\ref{70}), (\ref{100}) and (\ref{110})--(\ref{127}), one sees that 
\begin{equation}
\Pi \left( q\right) =\frac{\left\langle N\left( q\right) \right\rangle }N=%
\frac{p\left( q\right) x_s^q}{q!\Phi \left( x_s\right) }\,.  \label{130}
\end{equation}
This is valid for both the canonical and grand canonical ensembles. Note 
that Eq. (\ref{130}) may be also derived directly from the evolution
equation for the degree distribution (see Appendix \ref{evolution}).
Equations (\ref{120}) or (\ref{127}), and (\ref{130}) fix the one-to-one
correspondence between the degree distribution $\Pi \left( q\right) $, which
determines the microcanonical ensemble, and the set of parameters $\left( 
\bar{q},\left\{ f\left( q\right) \right\} \right) $ or, equivalently, $%
\left( \bar{q},\left\{ p\left( q\right) \right\} \right) $ (see Eq. (\ref{60}%
)). From Eqs. (\ref{60}) and (\ref{130}), it follows $\Pi \left( q+1\right)
/\Pi \left( q\right) =f\left( q\right) x_s/\left( q+1\right) $. Then one can
correspond the microcanonical ensemble which is described by a degree
distribution $\Pi \left( q\right) $ with the canonical and grand canonical
ensembles characterized by (i) 
\begin{equation}
f\left( q\right) =\left( q+1\right) \frac{\Pi \left( q+1\right) }{\Pi \left(
q\right) }
\,,  
\label{140}
\end{equation}
($f\left( q\right) $ is defined up to an arbitrary multiple), and (ii) by $%
\bar{q}=\sum_qq\Pi \left( q\right) $ for the canonical ensemble, or $\lambda
=\bar{q}$ for the grand canonical one.

\section{Generalization to the case of directed graphs}
\label{direct}

A microcanonical ensemble of directed graphs is characterized by a 
distribution function $\Pi \left( r,s\right) $, which is the probability 
that a randomly chosen vertex has in-degree $r$ and out-degree $s$. More
precisely, one must define non-negative integers $N_n\left( r,s\right) $ with
the following properties: $0<N_n=\sum_{r,s}N_n\left( r,s\right) <\infty $, $%
N_n\rightarrow \infty $ and $N_n\left( r,s\right) /N_n\rightarrow \Pi \left(
r,s\right) $ as $n\rightarrow \infty $. For a directed graph, we also must
require, that total in- and out-degrees are equal: $\sum_{r,s}\left(
r-s\right) N_n\left( r,s\right) =0$. Then for each $n$, we introduce the
ensemble of directed graphs with $N_n\left( r,s\right) $ vertices of
in-degree $r$ and out-degree $s$, $N_n$ vertices in total, connected in all
possible ways. To each this graph $g$, we ascribe for a statistical weight 
equal to the number of possible ways to construct the graph $g$:  
\begin{equation}
P_{MC}\left( g\right) =\prod_{i=1}^Nr_i!s_i!\prod_{j,k=1}^N\frac 1{g_{jk}!}%
\,.  \label{150}
\end{equation}
The limit of such a sequence at $n\rightarrow \infty $ would be the
microcanonical ensemble with a given degree distribution $\Pi \left(
r,s\right) $.

Canonical and grand canonical ensembles may be introduced quite analogously
to what it had been done for undirected graphs. For example, the canonical
ensemble may be introduced by using the process of rewiring one end of an
edge. There are two differences from the undirected graph constructions: (i)
we introduce two generalized preferential attachment functions, $f_1$ for
rewiring the outgoing end of an edge, and $f_2$ for rewiring the incoming
end, and (ii) in general, these function depend on both the in- and
out-degrees of the destination vertex. Applying detailed balance conditions
for transitions between graphs $g_1$ and $g_2$ (edge $i\rightarrow j$
rewires to $i\rightarrow k$), and between graphs $g_3$ and $g_4$ (edge $%
i\leftarrow j$ rewires to $i\leftarrow k$), we obtain the relations: 
\begin{eqnarray}
&&g_{ik}^{(2)}f_1\left( r_j^{(2)},s_j^{(2)}\right) P_C\left( g_2\right)
=g_{ij}^{(1)}f_1\left( r_k^{(1)},s_k^{(1)}\right) P_C\left( g_1\right) \,, 
\nonumber \\
&&g_{ij}^{(2)}=g_{ij}^{(1)}-1\,,\;g_{ik}^{(2)}=g_{ik}^{(1)}+1\,,\;%
\;r_j^{(2)}=r_j^{(1)}-1\,,\;s_j^{(2)}=s_j^{(1)}\,,\;r_k^{(2)}=r_k^{(1)}+1\,,%
\;s_k^{(2)}=s_k^{(1)}\,;  \label{170} \\
&&g_{ki}^{(4)}f_2\left( r_j^{(4)},s_j^{(4)}\right) P_C\left( g_4\right)
=g_{ji}^{(3)}f_2\left( r_k^{(3)},s_k^{(3)}\right) P_C\left( g_3\right) \,, 
\nonumber \\
&&g_{ki}^{(4)}=g_{ki}^{(3)}+1\,,\;g_{ji}^{(4)}=g_{ji}^{(3)}-1\,,%
\;r_j^{(4)}=r_j^{(3)}\,,\;s_j^{(4)}=s_j^{(3)}-1\,,\;r_k^{(4)}=r_k^{(3)}\,,%
\;s_k^{(2)}=s_k^{(1)}+1\,.  \label{180}
\end{eqnarray}
One can look for the solution of the above equations in the form: 
\begin{equation}
P_C\left( g\right) =\prod_{i=1}^Np\left( r_i,s_i\right) \prod_{j,k=1}^N\chi
\left( g_{jk}\right) \,.  \label{190}
\end{equation}
Then we have: 
\begin{eqnarray}
p\left( r+1,s\right) &=&f_1\left( r,s\right) p\left( r,s\right) \,,\;p\left(
r,s+1\right) =f_2\left( r,s\right) p\left( r,s\right) \,;  \label{200} \\
\chi \left( g+1\right) &=&\frac{\chi \left( g\right) }{g+1}\,.  \label{210}
\end{eqnarray}
Applying subsequently Eqs. (\ref{200}) in different order, we have: 
\[
p\left( r+1,s+1\right) =f_1\left( r,s+1\right) f_2\left( r,s\right) p\left(
r,s\right) =f_2\left( r+1,s\right) f_1\left( r,s\right) p\left( r,s\right)
\,. 
\]
This means that the preferential linking functions $f_1$ and $f_2$ cannot be
chosen arbitrary but must satisfy the condition 
\begin{equation}
f_1\left( r,s\right) f_2\left( r,s+1\right) =f_1\left( r+1,s\right)
f_2\left( r,s\right) \,,  \label{160}
\end{equation}
which is actually a consequence of the detailed balance condition (\ref{25}).

The solution of Eqs. (\ref{200}) is constructed in the following way. Let us
consider a 2D square lattice. We associate $f_1\left( r,s\right) $ with each
horizontal bond connecting sites $\left( r,s\right) $ and $\left(
r+1,s\right) $, and associate $f_2\left( r,s\right) $ with the vertical
bond, connecting sites $\left( r,s\right) $ and $\left( r,s+1\right) $. Let $%
{\cal L}$ be some path connecting points $\left( 0,0\right) $ with $\left(
r,s\right) $. Then,  
\begin{equation}
p\left( r,s\right) =p\left( 0,0\right) \prod_{{\cal L}}f_\alpha ^d\left(
\rho ,\sigma \right) \,.  \label{220}
\end{equation}
Here $\left( \rho ,\sigma \right) $ are coordinates of points along the path 
${\cal L}$, $\alpha =1\left( 2\right) $ for the horizontal (vertical) direction,
and $d=+1\left( -1\right) $ if the bond is passed in its positive (negative)
direction. The condition (\ref{160}) ensures the independence of the product
in Eq. (\ref{220}) of the path ${\cal L}$. In fact, this is the condition of
the potentiality (zero vorticity) of the vector field $\ln f_\alpha $
defined on the square lattice. Then $\ln p$ is a potential for this field,
that is $\ln f_\alpha $ is a lattice gradient of $\ln p$ (see Eq. (\ref{200}%
)). The arbitrary multiple $p\left( 0,0\right) $ may be set, e.g., to $1$. Solution
of Eq. (\ref{210}) is simple: 
\begin{equation}
\chi \left( g\right) =\frac 1{g!}\,.  \label{230}
\end{equation}

The grand canonical ensemble may be constructed quite analogously to what it had
been done for undirected graphs. Two opposite processes are introduced: one
is of edge creation, at a rate $f_2\left( r_i,s_i\right) f_1\left(
r_j,s_j\right) $ for the edge, going from the vertex $i$ to $j$, the other
is of the edge removal, at a rate $\lambda N$. Again, $f_1$ and $f_2$ must
satisfy the condition (\ref{160}) to ensure the equilibrium character of a
stationary state. The statistical weight of a graph is given by the
expression  
\begin{equation}
P_{GC}\left( g\right) =e^{-\lambda L\left( g\right) }\prod_{i=1}^Np\left(
r_i,s_i\right) \prod_{j,k=1}^N\frac 1{g_{jk}!}\,.  \label{240}
\end{equation}

The derivation of the integral representation of the partition function is
quite similar to that for undirected graphs, but Feynman's diagrams with
directed lines now are generated by complex fields. We introduce a complex
scalar field $x$, and write the action: 
\begin{eqnarray}
S\left( x,x^{*}\right) &=&-\Lambda \left| x\right| ^2-\varkappa \Phi \left(
x,x^{*}\right) \,,  \label{250} \\
\Phi \left( x,x^{*}\right) &=&\sum_{r,s=0}^\infty \frac{p\left( r,s\right) }{%
r!s!}x^r\left( x^{*}\right) ^s\,.  \label{260}
\end{eqnarray}
Then the following generating functional of this field theory will produce
all possible graphs with any number of vertices as its Feynman diagrams.
Their contributions are the same as statistical weights of graphs in the
grand canonical ensemble (\ref{240}) with $\lambda =\Lambda /N$, except the
additional multiples $\left( -\varkappa \right) ^N/N!$. Therefore, one can
write: 
\begin{equation}
Z\left( \varkappa ,\Lambda ,\left\{ p\left( r,s\right) \right\} \right) =%
\frac \Lambda \pi \int dxdx^{*}\,\exp S\left( x,x^{*}\right)
=\sum_{N=0}^\infty \frac{\left( -\varkappa \right) ^N}{N!}Z_{GC}\left(
N,\Lambda /N,\left\{ p\left( r,s\right) \right\} \right) \,,  \label{270}
\end{equation}
where the integration is over the entire complex plane (compare with Eq. (%
\ref{102})). Therefore,  
\begin{equation}
Z_{GC}\left( N,\lambda ,\left\{ p\left( r,s\right) \right\} \right) =\frac{%
N\lambda }\pi \int dxdx^{*}\,\exp \left( -N\lambda \left| x\right| ^2\right)
\left[ \Phi \left( x,x^{*}\right) \right] ^N\,.  \label{280}
\end{equation}
In Eqs. (\ref{270}), (\ref{280}) one should treat $x$ and $x^{*}$ as {\em %
independent} integration variables when actually calculating the integrals.
The partition function of the canonical ensemble is given by  
\begin{equation}
Z_C\left( N,L,\left\{ p\left( r,s\right) \right\} \right) =\frac{L!}{N^L}%
\oint_{C_1}\frac{dx}{2\pi i}\oint_{C_2}\frac{dy}{2\pi i}\left( xy\right)
^{-L-1}\left[ \Phi \left( x,y\right) \right] ^N\,,  \label{290}
\end{equation}
where the integration contours $C_{1,2}$ encircle points $x=0$, $y=0$,
respectively. The derivation of Eq. (\ref{290}) is quite similar to that of
Eq. (\ref{107}) for undirected graphs.

Again, in the thermodynamic limit, $N\rightarrow \infty $, $L\rightarrow
\infty $, $2L/N\rightarrow \bar{q}$, one can use a saddle point
approximation, which gives 
\begin{equation}
Z_C\left( N,L,\left\{ p\left( r,s\right) \right\} \right) \rightarrow \left( 
\frac{\bar{q}}{ex_sy_s}\right) ^L\left[ \Phi \left( x_s,y_s\right) \right]
^N\,,  \label{300}
\end{equation}
where $x_s$ and $y_s$ are defined from the stationary point equations: 
\begin{equation}
\bar{q}=x_s\frac{\partial \ln \Phi \left( x_s,y_s\right) }{\partial x_s}=y_s%
\frac{\partial \ln \Phi \left( x_s,y_s\right) }{\partial y_s}\,.  \label{310}
\end{equation}
For the grand canonical ensemble, we have  
\begin{equation}
Z_{GC}\left( N,\lambda ,\left\{ p\left( r,s\right) \right\} \right)
\rightarrow \exp \left( -N\lambda x_sy_s\right) \left[ \Phi \left(
x_s,y_s\right) \right] ^N\,,  \label{315}
\end{equation}
where the saddle point coordinates $x_s$ and $y_s$ are determined from the
equations: 
\begin{equation}
\lambda y_s=\frac{\partial \ln \Phi \left( x_s,y_s\right) }{\partial x_s}%
\,,\;\lambda x_s=\frac{\partial \ln \Phi \left( x_s,y_s\right) }{\partial y_s%
}\,.  \label{317}
\end{equation}

The degree distribution both for the canonical and grand canonical
ensembles is defined by  
\begin{equation}
\Pi \left( r,s\right) =\frac{\left\langle N\left( r,s\right) \right\rangle }N%
=\frac{\delta \ln Z\left( \left\{ p\left( u,v\right) \right\} \right) }{%
N\delta \ln p\left( r,s\right) }\rightarrow \frac{p\left( r,s\right) }{r!s!}%
x_s^ry_s^s\,,  \label{320}
\end{equation}
where the last relation is valid in the thermodynamic limit. Eqs. (\ref{320}%
), (\ref{310}) and (\ref{220}) establish correspondence between the
microcanonical ensemble with the degree distribution $\Pi \left( r,s\right) $
and the canonical one, characterized by the preferential linking functions $%
f_{1,2}$ and the mean vertex degree $\bar{q}$. The parameters of the
canonical and grand canonical ensembles with the same degree distribution
are related as  
\begin{equation}
\bar{q}=\lambda x_sy_s\,.  \label{330}
\end{equation}
This relation follows from Eqs. (\ref{310}) and (\ref{317}).

Again, as it was for undirected graphs, the canonical ensemble does exist
for every $p\left( r,s\right) $, provided that power series (\ref{260}) has
finite radii of convergence on both $x$ and $y$. Conditions for the
existence of the grand canonical ensemble are essentially more strict: the
integral in Eq. (\ref{280}) must be well defined and convergent.

\section{Fat-tailed degree distributions}
\label{fat}

In this section we shall consider in detail properties of the canonical ensembles of
graphs, which arise if a preference function $f(q)$ grows rapidly enough.

For brevity, we focus on the undirected graphs. The generalization to
the ensembles of directed graphs is straightforward.

As one can see from Eq. (\ref{90}), the grand canonical ensemble does not
exist in two cases. In the first case the integral, representing the
partition function, diverges, since the function $\Phi \left( x\right) $
grows fast enough at $x\rightarrow \pm \infty $. In the second case, the
integral is not determined, because $\Phi \left( x\right) $ has a singularity
on the real axis. In both the situations we have degree
distributions, which decay relatively slowly as $q\rightarrow \infty $%
. Let us begin with the case, when $\Phi \left( x\right) $ has no
singularities, but $\ln \Phi \left( x\right) $ grows faster than $x^2$ as $%
\left| x\right| \rightarrow \infty $.

Using Eqs. (\ref{100}) and (\ref{130}), one can write the following
relation: 
\begin{equation}
\frac{\Phi \left( x\right) }{\Phi \left( x_s\right) }=\sum_{q=0}^\infty \Pi
\left( q\right) \left( \frac x{x_s}\right) ^q\,,  \label{340}
\end{equation}
that is $\Phi \left( x\right) $ is expressed in terms of the $Z$-transform of $%
\Pi \left( q\right) $. Using the formula for the inverse of $Z$-transform, we
obtain 
\begin{equation}
\Pi \left( q\right) =\oint \frac{dx}{2\pi ix}\left( \frac{x_s}x\right) ^q%
\frac{\Phi \left( x\right) }{\Phi \left( x_s\right) }\,.  \label{350}
\end{equation}
For finding the relation between the asymptotic behaviours of $\Phi \left(
x\right) $ and $\Pi \left( q\right) $, let us use a saddle point
approximation in Eq. (\ref{350}). It is convenient to set $\Phi \left(
x\right) =\exp \phi \left( x\right) $. The equation for the saddle point $%
x_a $ is $q=x_a\phi ^{\prime }\left( x_a\right) $. Then the asymptotic
expression for $\Pi \left( q\right) $ is 
\begin{equation}
\Pi \left( q\right) \rightarrow \left\{ 2\pi x_a\left[ x_a\phi ^{\prime
\prime }\left( x_a\right) -\phi ^{\prime }\left( x_a\right) \right] \right\}
^{-1/2}\left( \frac{x_s}{x_a}\right) ^q\exp \left[ \phi \left( x_a\right)
-\phi \left( x_s\right) \right]  \label{360}
\end{equation}
The integral for the grand canonical partition function in Eq. (\ref{90}) is
divergent, if $\phi \left( x\right) $ grows as $x^2$ 
or faster at $x\rightarrow \infty $. Assume that $\phi \left( x\right) \rightarrow Ax^\mu $ as $%
x\rightarrow \infty $. Then the saddle point equation is $q=A\mu x^\mu $,
and $x_a\rightarrow \left( q/A\mu \right) ^{1/\mu }$. Omitting irrelevant
multiples, we have from Eq. (\ref{360}): 
\begin{equation}
\Pi \left( q\right) \sim \left( 2\pi q\right) ^{-1/2}\left( \frac{\bar{q}}q%
e\right) ^{q/\mu }\sim \left[ \Gamma \left( q\right) \right] ^{-1/\mu }\,.
\label{370}
\end{equation}
Thus, if the degree distribution $\Pi \left( q\right) $ decays slower than $%
\left[ \Gamma \left( q\right) \right] ^{-1/2}$ as $q\rightarrow \infty $,
then the partition function of the corresponding grand canonical ensemble
diverges.

The reason for this divergence is that we have admitted the existence of
non-Mayer's graphs. Indeed, let us choose some pair of vertices in a graph.
Let us add more and more edges connecting this pair. The statistical weight
of the graph with $\nu $ edges between this pair contains multiple $p\left(
q_i\right) p\left( q_j\right) /\nu _{ij}!$, where $q_i=\nu _{ij}+{\rm const}$%
, $q_j=\nu _{ij}+{\rm const}$. One can easily conclude that the statistical
weights approach $0$ as $\nu \rightarrow \infty $ only if $p^2\left(
q\right) /q!\rightarrow 0$, or $\sqrt{q!}\Pi \left( q\right) \rightarrow 0$.
The same result may be attained if we consider a sequence of graphs 
obtained by subsequent addition of closed loops to a chosen vertex. This
result may also be presented in the different way: if the preference 
function $f\left( q\right) =p\left( q+1\right) /p\left( q\right) $ grows 
faster than $\sqrt{q}$ at large $q$, then the partition function of the grand
canonical ensemble diverges.

The radius of convergence of the series expansion (\ref{100}) is  
\begin{equation}
R_c=\lim_{q\rightarrow \infty }\frac{\left( q+1\right) p\left( q\right) }{%
p\left( q+1\right) }=\lim_{q\rightarrow \infty }\frac{q+1}{f\left( q\right) }%
\,.  \label{3710}
\end{equation}
If $f\left( q\right) $ grows as $q\rightarrow \infty $ slower than a linear
function, then $R_c=\infty $. In this case $\Phi \left( x\right) $ has no
singularities at all. This means that (i) the partition function of the
canonical ensemble may be expressed in the form (\ref{107}), and (ii) in the
thermodynamic limit, one can use for this function its saddle point
expression, Eqs. (\ref{110}) and (\ref{120}). If $f\left( q\right) $ grows 
faster than a linear function, then $R_c=0$. This means that although the
canonical ensemble exists (the canonical ensemble always exists, because it is represented by a 
{\em finite} set of graphs at any finite $N$ and $L$), its partition
function can not be written in the form of the integral representation (\ref{107}).
Actually, this means the absence of any meaningful thermodynamic limit. 

The
interesting case is $0<R_c<\infty $. In this case, the partition function of the
canonical ensemble can be expressed as an integral, but the saddle point
expression for this integral may not be longer valid. 
The saddle point expression is not valid at a 
large enough number of edges in the network, when the saddle point 
approaches the position of singularity. We show that in this situation, ``fat-tailed'' degree
distributions, i.e. ones decreasing slower than an exponent, may arise.

Without any lack of generality, one can set $R_c=1$ in this case. This is
equivalent to $f\left( q\right) =q+o\left( q\right) $ as $q\rightarrow
\infty $. In Eq. (\ref{120}) its right hand side is a monotonously
increasing function of $x_s$. This means that as $\bar{q}=2L/N$ grows, $x_s$
grows too. As $x_s<R_c=1$, the degree distribution contains
exponentially decaying multiple $x_s^q$. There are two possibilities
depending on the character of the singularity of $\Phi \left( x\right) $ at $%
x=1$: either $\Phi ^{\prime }\left( x_s\right) \rightarrow \infty $, or it
approaches some finite value as $x_s\rightarrow \infty $. In the former
case, again there are two possibilities: either $\lim_{x\rightarrow 1}\Phi
\left( x\right) $ is finite, or this limit is infinite. If $\Phi \left( 1\right) $
is finite (but $\Phi ^{\prime }\left( 1\right) $ is infinite), then the
degree distribution approaches some limiting form as $\bar{q}%
\rightarrow \infty $, and the first moment of this limiting distribution
diverges. This means that such a degree distribution can not be realized in
any canonical ensemble with a finite number of edges per vertex. To
construct networks with such a distribution, one has to change the
conditions of the thermodynamic limit transition in the canonical ensemble,
assuming $N\rightarrow \infty $, $L\rightarrow \infty $ {\em and} $\bar{q}%
=2L/N\rightarrow \infty $, instead of keeping $\bar{q}$ fixed. Another way
is to use a microcanonical ensemble. If $\Phi \left( 1\right) $ is infinite,
no normalizable degree distribution without an exponential cut-off is possible.

Now, let us consider the case $\Phi ^{\prime }\left( 1\right) <\infty $. The
degree distribution becomes ``fat-tailed'' when $x_s=1$, which takes place
when $\bar{q}=\bar{q}_c=\Phi ^{\prime }\left( 1\right) /\Phi \left( 1\right) 
$. If $\bar{q}>q_c$, the saddle point equation (\ref{120}) has no solution $%
0<x_s<1$. In this case, in the thermodynamic limit, the partition function 
remains the same up to a preexponential factor as for $\bar{q}=q_c$.
Indeed, let us rewrite Eq. (\ref{107}) as  
\begin{equation}
Z_C\left( N,L,\left\{ p\left( q\right) \right\} \right) =N^{-L}\left(
2L-1\right) !!\oint_c\frac{dx}{2\pi ix}\left[ x^{-\bar{q}}\Phi \left(
x\right) \right] ^N.  \label{3720}
\end{equation}
To calculate a large $N$ asymptotics one has to deform the integration contour
into the steepest descent one, intercepting the real axis at the point, where $%
x^{-\bar{q}}\Phi \left( x\right) $ is minimal within the interval $\left(
0,1\right) $, and going along the line of the constant (i.e. zero) imaginary
part. If $\bar{q}<q_c$, this is a usual saddle-point contour, crossing the
real axis perpendicularly at some point $x_s<1$. If $\bar{q}>q_c$, this
contour consists of two complex conjugate parts meeting always at $x=1$. As $%
\bar{q}$ grows, the point, where the integrand is maximal, $x_s=1$, does
not move. The only change is that the two branches of the contour become 
closer and closer  
to the real axis in the vicinity of $x=1$ at $x>1$. 
But it is $x_s$ and $\Phi \left( x_s\right) $ 
that determine the
value of the main (extensive) contribution to the logarithm of the partition
function: $\ln Z_C=-L\ln \left( \bar{q}x_s\right) +N\ln \Phi \left(
x_s\right) +o\left( N\right) $. The extensive part of the ``free energy'' $%
-\ln Z$ does not depend on $\bar{q}$ as $\bar{q}>q_c$. 
So, the degree
distribution $N^{-1}\delta \ln Z/\delta \ln p\left( q\right) $ (see
Eqs. (\ref{17}) and (\ref{70})) remains equal to its critical point value $%
\Pi _c\left( q\right) $. Consequently, the finite fraction of edges, $\bar{q}%
/\bar{q}_c-1$, is attached to an infinitely small fraction of vertices,
forming a ``condensate'', quite analogous to the one in the backgammon model 
\cite{bbj99,bjj02}.

A specific form of the degree distribution at the critical point depends on
the behaviour of this difference $f\left( q\right) -q=o\left( q\right) $ as $%
q\to \infty $. For example, for the so called ``scale-free'' distributions, $%
\Pi \left( q\right) \propto q^{-\gamma }$ as $q\to \infty $ ($\gamma >2$),
we obtain from Eq. (\ref{140}):
\begin{equation}
f\left( q\right) =q+1-\gamma +{\cal O}\left( q^{-1}\right)  \label{1500}
\end{equation}
at the critical point. For the same $f\left( q\right) $ but for lower
average degrees $\bar{q}<\bar{q}_c$, we have $\Pi \left( q\right) \propto
x_s^{-q}q^{-\gamma }$ at large $q$. So, the state with a power-law degree
distribution is marginal for the phase without the condensate of edges \cite
{remark3}. (A ``scale-free'' state as a line between ``generic'' and
``crumpled'' phases on the phase diagram of trees was found in Ref. \cite
{bck01}, see also the condensation transition in the backgammon model \cite
{bbj99}.) Our analysis has shown that the condensation takes place above $%
\bar{q}_c$ \cite{remark5}. Furthermore, the fat-tailed degree distribution
is present also in the condensed phase. The problem of the condensed phase
is more complex for the ensembles of Mayer's graphs, that is the ones
without tadpoles and melons. The nature of condensation transition in such
ensembles will be discussed elsewhere.

\section{Conclusions}
\label{concl}

Thus, we have developed the consistent description of random networks 
in the framework of classical statistical mechanics. 
Using the traditional formalism of statistical mechanics, we have
constructed a set of equilibrium statistical ensembles of uncorrelated 
random networks and have found their partition functions and main
characteristics. 
We have proposed a set of natural dynamical procedures, which generate 
equilibrium networks as a limiting state of the evolution, and 
have established a one-to-one correspondence between rules of these ergodic 
procedures and equilibrium ensembles of networks. 
This program has been realized both for directed and undirected networks. 
 
We have shown that a ``scale-free'' state 
(and fat-tailed degree distributions) in equilibrium uncorrelated 
networks {\em without condensation of edges on vertices} may exist only in a single marginal point. 
So, it is rather an exception \cite{remark4}. This differs crucially
from the situation for growing networks. The latter, while growing, may 
self-organize into scale-free structures in a wide range of parameters
without any condensation. In summary, we have developed a statistical physics
approach to equilibrium random networks. 

S.N.D. thanks PRAXIS XXI (Portugal) for a research grant PRAXIS
XXI/BCC/16418/98. S.N.D. and J.F.F.M. were partially supported by the
project POCTI/99/FIS/33141. A.N.S. acknowledges the NATO program OUTREACH
for support. We also thank V.V.~Bryksin, A.V.~Goltsev, A.~Krzywicki, and
F.~Slanina for useful discussions.

\appendix

\section{Statistical weights for the microcanonical ensemble}
\label{combinatorics}

Initially, we have $N$ vertices (``hedgehogs'') with $q_i$ edges (``halves'' of edges, speaking more precisely) protruding
from an $i$-th one. Here we count the number of ways of connecting them in pairs 
to obtain a given graph with a given number of edges $g_{ij}$ between
vertices $i$ and $j$. 
For vertices with unit-length loops
(``tadpoles'') we set $g_{ii}$ to be equal to twice the number of such
loops. The number of ways to choose $g_{i1},g_{i2},\dots g_{iN}$ edges from $%
q_i=g_{i1}+g_{i2}+\cdots +g_{iN}$ ones, attached to the $i$-th vertex is 
\begin{equation}
\frac{q_i!}{g_{i1}!g_{i2}!\cdots g_{iN}!}\,.  \label{380}
\end{equation} 

Then we have to connect in pairs $g_{ij}$ dangling edges, attached to the $i$-th
vertex, and $g_{ji}=g_{ij}$ edges attached to the $j\ne i$ vertex. This can be done by $%
g_{ij}!$ different ways. Also, the number of ways to join $g_{ii}$ dangling
edges in pairs to form $g_{ii}/2$ closed loops is $\left( g_{ii}-1\right)
\left( g_{ii}-3\right) \cdots 1=\left( g_{ii}-1\right) !!$. Finally,
combining together $N$ multiples (\ref{380}) for each vertex, $N\left(
N-1\right) /2$ multiples $g_{ij}!$ for each pair of vertices, multiples $%
\left( g_{ii}-1\right) !!$ for each vertex, containing unit-length loops,
and taking into account that $\left( g_{ii}-1\right) !!/g_{ii}!=1/g_{ii}!!$,
we arrive at Eq. (\ref{35}).

\section{Integral representation of the partition function for canonical
ensemble}
\label{diagrams}

The partition function of the canonical ensemble is  
\begin{equation}
Z_C\left( N,L\right) =N^{-L}\sum_{g\in \Omega \left( N,L\right)
}\prod_{i=1}^N\frac{p\left( q_i\right) }{g_{ii}!!}\prod_{j<k=1}^N\frac 1{%
g_{jk}!}\,,  \label{390}
\end{equation}
where the set $\Omega \left( N,L\right) $ is a set of $N^2$ non-negative
integers $g_{ij}\ge 0$ with the following properties: (i) $g_{ii}$ are even,
(ii) $g_{ij}=g_{ji}$, and (iii) 
\begin{equation}
\frac 12\sum_{i,j=1}^Ng_{ij}=\sum_{i=1}^N\frac{g_{ii}}2%
+\sum_{i>j=1}^Ng_{ij}=L\,.  \label{400}
\end{equation}
So, $N\left( N+1\right) /2$ variables $g_{ij}$, $i\ge j$, are subjected to
the restriction (\ref{400}). Introducing  
\begin{equation}
\Phi \left( x\right) =\sum_{q=0}^\infty \frac{p\left( q\right) }{q!}x^q\,,
\label{410}
\end{equation}
one can write Eq. (\ref{390}) as
\begin{eqnarray}
&&Z_C\left( N,L\right)   
\nonumber 
\\[5pt]
&=&\left. N^{-L}\sum_{\left\{ g\right\} \in \Omega \left( N,L\right)
}\prod_{i=1}^N\left[ \left( \frac{g_{ii}}2\right) !2^{g_{ii}/2}\right]
^{-1}\left( \frac{\partial ^2}{\partial x_i^2}\right)
^{g_{ii}/2}\prod_{j<k=1}^N\left( g_{jk}!\right) ^{-1}\frac{\partial ^2}{%
\partial x_j\partial x_k}\prod_{l=1}^N\Phi \left( x_l\right) \right|
_{x_1=\cdots =x_N=0}  
\nonumber 
\\[5pt]
&=&\left. \frac{\left( 2N\right) ^{-L}}{L!}\left( \sum_{i=1}^N\frac{\partial
^2}{\partial x_i^2}+2\sum_{i>j=1}^N\frac{\partial ^2}{\partial x_i\partial
x_j}\right) ^L\prod_{l=1}^N\Phi \left( x_l\right) \right| _{x_1=\cdots
=x_N=0}  
\nonumber 
\\[5pt]
&=&\left. \frac{\left( 2N\right) ^{-L}}{L!}\left( \sum_{i=1}^N\frac \partial
{\partial x_i}\right) ^{2L}\prod_{l=1}^N\Phi \left( x_l\right) \right|
_{x_1=\cdots =x_N=0}
\,,  
\label{420}
\end{eqnarray}
where the relation (\ref{400}) was used. If we pass from $x_1,\dots ,x_N$ to
a new set of variables: $x=\left( x_1+\cdots +x_N\right) /N$, and difference
variables $y_i=x_i-x_{i+1}$, $i=1,\dots ,N-1$, we have 
\[
\frac \partial {\partial x}=\sum_{i=1}^N\frac{\partial x_i}{\partial x}\frac %
\partial {\partial x_i}=\sum_{i=1}^N\frac \partial {\partial x_i}\,.
\]
Then 
\begin{equation}
Z_C\left( N,L\right) =\left. \frac{\left( 2N\right) ^{-L}}{L!}\frac{\partial
^{2L}}{\partial x^{2L}}\left[ \Phi \left( x\right) \right] ^N\right|
_{x=0}\,.  \label{430}
\end{equation}
Finally, one can write  
\begin{equation}
Z_C\left( N,L\right) =\frac{\left( 2N\right) ^{-L}}{L!}\left( 2L\right)
!\oint_c\frac{dx}{2\pi i}\,x^{-2L-1}\left[ \Phi \left( x\right) \right] ^N\,,
\label{440}
\end{equation}
which is exactly Eq. (\ref{107}). The contour $c$ encircles the point $x=0$.

\section{Evolution equation for the degree distribution}
\label{evolution}

Here we present a simplified derivation of the evolution equation for the
degree distribution $\Pi \left( q,t\right) $. For example, we consider a
network with the rewiring of edges according to the rules formulated in
Section \ref{undirect} for the canonical ensemble (for the grand canonical
ensemble, the procedure is essentially the same). The total number of
vertices, $N$, and of edges, $L$, are fixed. With some probability $n$ per
unit time, a randomly chosen end of a randomly chosen edge is rewired to
some vertex of the graph. This vertex is chosen from vertices of the graph
with probability proportional to a given function $f\left( q_i\right) $ of
the degree $q_i$ of the vertex.

Then, the probability that a vertex $i$ {\em receives} a new edge per time $%
dt$ is  
\begin{equation}
\frac{f\left( q_i\right) n\,dt}{\sum_jf\left( q_j\right) }\rightarrow \frac{%
nf\left( q_i\right) }{N\overline{f\left( q\right) }}\,.  \label{450}
\end{equation}
In Eq. (\ref{450}) its left-hand side may be replaced with the right-hand
side in the thermodynamic limit $N\rightarrow \infty $ (self-averaging) if
the fluctuations of $N\left( q\right) $ can be neglected. This is true for
the equilibrium state (see Eq. (\ref{70})), and so here we restrict
ourselves to the nonequilibrium states that fulfill this condition. Then, the
infinitesimal change of the degree distribution due to the rewiring of
edges {\em to} a chosen vertex is
\begin{equation}
\left[ d\Pi \left( q,t\right) \right] _{{\rm to}}=\frac n{N\overline{f\left(
q\right) }}\left[ f\left( q-1\right) \Pi \left( q-1,t\right) -f\left(
q\right) \Pi \left( q,t\right) \right] dt\,.  \label{460}
\end{equation}

Also, we must take into account that the vertex may loose one of its $q_i$
edges, which will be rewired to another vertex. The probability $nq_idt/L$
that one of these edges is chosen for rewiring per time $dt$, must be
multiplied by $1/2$. This is the probability that of the two ends of the
edge, the one attached to the $i$-th vertex, is chosen. Thus, the change of
the degree distribution due to rewiring of edges {\em from} a vertex is
\begin{equation}
\left[ d\Pi \left( q,t\right) \right] _{{\rm from}}=\frac n{N\overline{q}}%
\left[ \left( q+1\right) \Pi \left( q+1,t\right) -q\Pi \left( q,t\right)
\right] 
\,,  
\label{470}
\end{equation}
where $\bar{q}=2L/N$ is the average vertex degree. Combining Eqs. (\ref{460}%
) and (\ref{470}) we arrive at the evolution equation  
\begin{equation}
\frac Nn\frac{\partial \Pi \left( q,t\right) }{\partial t}=\frac 1{\overline{%
f\left( q\right) }}\left[ f\left( q-1\right) \Pi \left( q-1,t\right)
-f\left( q\right) \Pi \left( q,t\right) \right] +\frac 1{\bar{q}}\left[
\left( q+1\right) \Pi \left( q+1,t\right) -q\Pi \left( q,t\right) \right] 
\,.
\label{480}
\end{equation}

Looking for the stationary solution of Eq. (\ref{480}), one can easily find
its first integral: 
\begin{equation}
\frac{q+1}{\bar{q}}\Pi \left( q+1\right) -\frac{f\left( q\right) }{\overline{%
f\left( q\right) }}\Pi \left( q\right) ={\rm const\,.}  \label{490}
\end{equation}
One must set ${\rm const}=0$ in Eq. (\ref{490}), because $\Pi \left(
q\right) =0$ at $q<0$. Then we have  
\begin{equation}
\Pi \left( q+1\right) =x_s\frac{f\left( q\right) }{q+1}\Pi \left( q\right)
\,,  \label{500}
\end{equation}
where we have introduced $x_s=\bar{q}/\overline{f\left( q\right) }$. The
solution of Eq. (\ref{500}) is  
\begin{equation}
\Pi \left( q\right) =C\frac{p\left( q\right) }{q!}x_s^q\,.  \label{510}
\end{equation}
Here $C$ and $x_s$ must be determined from the normalization condition and
from the equality of the mean degree to a given value $\bar{q}=2L/N$: 
\begin{equation}
\sum_{q=0}^\infty \Pi \left( q\right) =1\,,\;\sum_{q=0}^\infty q\Pi \left(
q\right) =\bar{q}\,.  \label{520}
\end{equation}

\noindent
$^{*}$ E-mail address: sdorogov@fc.up.pt \newline
$^{\dagger }$ E-mail address: jfmendes@fis.ua.pt \newline
$^{\ddagger }$ E-mail address: samukhin@fc.up.pt



\begin{references}
\bibitem{remark0}  Notice that in graph theory, ``random graphs'' are
usually networks with a Poisson distribution of connections.
\bibitem{remark0a}  To approach thermodynamic limit of the microcanonical
ensemble, one may consider a sequence of microcanonical ensembles: $%
\sum_qN_n\left( q\right) =N_n\to \infty $, $N_n\left( q\right) /N_n\to \Pi
\left( q\right) $ as $n\to \infty $. Here $\Pi (q)$ is the degree
distribution.
\bibitem{remark0b}  In the thermodynamic limit, this is equivalent to
rewiring both the ends of a randomly chosen edge to vertices $i$ and $k$
at the rate $f\left( q_i\right) f\left( q_k\right) $.
\bibitem{remark3}  In contrast, in growing networks, power-law degree
distributions are realized in the case of any linear preference function $f\left(
q\right) $. In growing networks, if $f\left( q\right) \sim q^\delta $, $\delta <1$, the
distribution has a size-independent cutoff; if $\delta >1$, a finite
fraction of edges are attached to a single vertex \cite{krl00}.
\bibitem{remark4}  Main analytical results on percolation and disease spread
on scale-free networks were obtained just for these exceptional equilibrium
uncorrelated scale-free nets without condensate.
\end{references}
\end{document}